\documentclass[article,preprint,groupedaddress,11pt]{revtex4}
\usepackage{epsfig}
\usepackage{graphicx}
\usepackage{amssymb}

\begin{document}
\title{A proof of the weak gravity conjecture}
\author{Shahar Hod}
\affiliation{The Ruppin Academic Center, Emeq Hefer 40250, Israel}
\affiliation{ }
\affiliation{The Hadassah Institute, Jerusalem 91010, Israel}
\date{\today}
\centerline {\it This essay is awarded 4th Prize in the 2017 Essay
Competition of the Gravity Research Foundation}

\begin{abstract}
\ \ \ The weak gravity conjecture suggests that, in a
self-consistent theory of quantum gravity, the strength of gravity
is bounded from above by the strengths of the various gauge forces
in the theory. In particular, this intriguing conjecture asserts
that in a theory describing a U(1) gauge field coupled consistently
to gravity, there must exist a particle whose proper mass is bounded
(in Planck units) by its charge: $m/m_{\text{P}}<q$. This beautiful
and remarkably compact conjecture has attracted the attention of
physicists and mathematicians over the last decade. It should be
emphasized, however, that despite the fact that there are numerous
examples from field theory and string theory that support the
conjecture, we still lack a general proof of its validity. In the
present Letter we prove that the weak gravity conjecture (and, in
particular, the mass-charge upper bound $m/m_{\text{P}}<q$) can be
inferred directly from Bekenstein's generalized second law of
thermodynamics, a law which is widely believed to reflect a
fundamental aspect of the elusive theory of quantum gravity.
\newline
\newline
Email: shaharhod@gmail.com
\end{abstract}
\bigskip
\maketitle

It is widely believed that string theory may provide a
self-consistent description of the elusive theory of quantum
gravity. However, the predictive power of the theory seems to be
restricted by its permissive nature \cite{wgc}: there are simply too
many semi-classically consistent theories of gravity that arise in
string theory.

In order to distinguish the landscape of consistent theories of
gravity from the swampland of low-energy effective theories which
cannot be completed to a full theory of quantum gravity,
Arkani-Hamed, Motl, Nicolis, and Vafa \cite{wgc} have proposed the
intriguing ``weak gravity conjecture" as a very simple and powerful
constraint on the possible gauge theories that could arise from a
self-consistent quantum theory of gravity. In its simplest form,
this highly interesting conjecture asserts that in a theory
describing a U(1) gauge field coupled consistently to gravity, there
must exist at least one state (particle) whose proper mass is
bounded from above by its charge:
\begin{equation}\label{Eq1}
m/m_{\text{P}}<q\  ,
\end{equation}
where $m_{\text{P}}\equiv \sqrt{\hbar c/G}$ is the fundamental
Planck mass \cite{Noteelc,Noteunit}.

As originally discussed in \cite{wgc}, a violation of the weak
gravity conjecture (\ref{Eq1}) in a theory describing a U(1) gauge
field coupled to gravity would imply the absolute stability of
extremal black holes in this theory. Since entropic arguments
suggest that stable black-hole remnants are pathological in a
quantum theory of gravity \cite{Beke,Sus,Bek1}, it has been asserted
in \cite{wgc} that the mass-charge relation (\ref{Eq1}) is mandatory
in any self-consistent theory of quantum gravity \cite{Notehe}.

The elegant and remarkably compact weak-gravity conjecture
(\ref{Eq1}) has attracted the attention of physicists and
mathematicians over the last decade (see
\cite{wr1,wr2,wr3,wr4,wr5,wr6,wr7} and references therein). However,
it is important to emphasize that despite the fact that there are
numerous examples from field theory and string theory that support
the conjecture \cite{wgc,wr1,wr2,wr3,wr4,wr5,wr6,wr7}, we still lack
a general proof of its validity.

One naturally wonders: where does the conjectured weak gravity bound
$m/q<1$ come from? It is not clear how to derive the suggested
mass-charge relation (\ref{Eq1}) directly from microscopic quantum
considerations. In particular, microscopic physics seems to afford
no special status to the dimensionless mass-to-charge ratio $m/q$ of
fundamental fields.

It should be emphasized that the weak gravity conjecture is expected
to characterize gauge field theories in a self-consistent quantum
theory of gravity \cite{wgc}. This fact suggests that a derivation
of the conjectured mass-charge relation (\ref{Eq1}), which quantify
the weak gravity principle, may require use of the yet unknown
quantum theory of gravity. This conclusion, if true, may seem as bad
news for our physical aspirations to provide a general proof of the
weak gravity bound $m/q<1$. But we need not lose heart -- it is
widely believed \cite{Foux} that Bekenstein's generalized second law
of thermodynamics \cite{Bek1,Notegsl}, and the closely related
concept of black-hole entropy (which combines together all three
fundamental constants of nature into one simple formula
$S_{\text{BH}}=k_{\text{B}}c^3A/4G\hbar$ \cite{Bek1,Noteaa}),
reflect a fundamental aspect of any self-consistent quantum theory
of gravity.

Interestingly, and most importantly for our analysis, it has been
explicitly shown in \cite{Hod1,Hod2} that the generalized second law
of thermodynamics \cite{Bek1} yields a fundamental lower bound on
the characteristic relaxation time $\tau$ of perturbed physical
systems. In particular, for a thermodynamic system of temperature
$T$, the universal relaxation bound can be expressed by the compact
time-times-temperature (TTT) relation \cite{Hod1,Hod2}
\begin{equation}\label{Eq2}
\tau\times T\geq 1/\pi\  .
\end{equation}

In the context of a self-consistent quantum theory of gravity, the
universal relaxation bound (\ref{Eq2}) asserts that black holes,
which are known to be characterized by a well defined
Bekenstein-Hawking temperature $T_{\text{BH}}$ \cite{Bek1}, must
have (at least) one exponentially decaying perturbation mode
$\Psi(r,t)=\psi(r)e^{-i\omega t}$ whose fundamental resonant
frequency $\omega_0$ is characterized by the relation
\cite{Noterelx,Noterec}
\begin{equation}\label{Eq3}
\Im\omega_0\leq \pi T_{\text{BH}}\  .
\end{equation}
Interestingly, it has been proved \cite{Hod1,Hod2} that
astrophysically realistic (spinning) Kerr black holes conform to the
universal relaxation bound (\ref{Eq2}). In particular, the
fundamental inequality (\ref{Eq3}) is saturated in the near-extremal
$T_{\text{BH}}\to0$ limit of rapidly-rotating Kerr black holes
\cite{Hod1,Hod2}.

We shall now show explicitly that the universal relaxation bound
(\ref{Eq2}), when applied to the characteristic resonant relaxation
spectrum of near-extremal charged Reissner-Nordstr\"om black holes,
yields in a remarkably simple way the fundamental weak gravity
(mass-charge) relation (\ref{Eq1}).

{\it The universal relaxation bound and the weak gravity
conjecture.---} The late-time relaxation dynamics of charged massive
scalar fields in the charged Reissner-Nordstr\"om black-hole
spacetime is known to be characterized by exponentially damped
oscillations \cite{Hodch1,Hodch2,Konf,Ric,Herc}. In particular, the
characteristic relaxation timescale $\tau_{\text{relax}}$ of the
perturbed black-hole spacetime is determined by the imaginary part
of the fundamental (least damped) resonant mode of the composed
black-hole-field system:
\begin{equation}\label{Eq4}
\tau_{\text{relax}}\equiv 1/\Im\omega_0\  .
\end{equation}
As we shall now prove, the intriguing mass-charge (weak gravity)
inequality (\ref{Eq1}) can be deduced directly from the universal
relaxation bound (\ref{Eq3}) as applied to the characteristic
complex relaxation spectrum of near-extremal charged
Reissner-Nordstr\"om black-hole spacetimes.

The physical properties of a linearized perturbation mode
\cite{Notecor,Noteom}
\begin{equation}\label{Eq5}
\Psi(t,r,\theta,\phi)=r^{-1}e^{im\phi}S_{lm}(\theta)\psi_{lm}(r;\omega)e^{-i\omega
t}\
\end{equation}
describing the dynamics of a scalar field $\Psi$ of proper mass
$\mu$ and charge coupling constant $q$ in a Reissner-Nordstr\"om
black-hole spacetime of mass $M$ and electric charge $Q$ are
determined by the Schr\"odinger-like ordinary differential equation
\cite{Hodch1,Hodch2,Konf,Ric,Herc,Noters}
\begin{equation}\label{Eq6}
{{d^2\psi}\over{dr^{*2}}}+V\psi=0\  ,
\end{equation}
where the effective radial potential which characterizes the
interaction of the charged massive field with the curved black-hole
spacetime is given by \cite{Hodch1,Hodch2,Konf,Ric,Herc}
\begin{equation}\label{Eq7}
V=V(r;M,Q,\omega,q,\mu,l)=\Big(\omega-{{qQ}\over{r}}\Big)^2-
\Big(1-{{2M}\over{r}}+{{Q^2}\over{r^2}}\Big)
\Big[\mu^2+{{l(l+1)}\over{r^2}}+{{2M}\over{r^3}}-{{2Q^2}\over{r^4}}\Big]\
.
\end{equation}
We are interested in resonant perturbation modes of the composed
black-hole-field system which are characterized by the physically
motivated boundary condition $\psi(r\to r_+)\sim
e^{-i(\omega-qQ/r_+)r^*}$ of purely ingoing waves at the horizon
$r_+=M+(M^2-Q^2)^{1/2}$ of the black hole. In addition, the
black-hole-field perturbation modes are characterized by the
boundary condition $\psi(r\to \infty)\sim
e^{i\sqrt{\omega^2-\mu^2}r}$ at spatial infinity \cite{Noteai}. The
Schr\"odinger-like differential equation (\ref{Eq6}) with the
effective radial potential (\ref{Eq7}) and the above stated
physically motivated boundary conditions determine the complex
resonant spectrum $\{\omega_n(M,Q,\mu,q,l)\}_{n=0}^{n=\infty}$ which
characterizes the relaxation dynamics of the composed
Reissner-Nordstr\"om-black-hole-charged-massive-scalar-field system.

Interestingly, as explicitly shown in \cite{Hodch1}, the fundamental
(least damped) resonances of this composed black-hole-field system
can be determined {\it analytically} in the regime \cite{Noterpm}
\begin{equation}\label{Eq8}
T_{\text{BH}}={{\hbar(r_+-r_-)}\over{4\pi r^2_+}}\to0\
\end{equation}
of near-extremal black holes. In particular, from equations
(\ref{Eq6})-(\ref{Eq7}) one finds the remarkably compact expression
\cite{Hodch1}
\begin{equation}\label{Eq9}
\Im\omega_n=2\pi T_{\text{BH}}(n+1/2+\Im\delta)\ \ \ ; \ \ \
n=0,1,2,...
\end{equation}
for the imaginary parts of the composed black-hole-field resonances
in the near-extremal regime (\ref{Eq8}), where the dimensionless
field parameter $\delta$ is given by \cite{Hodch1}
\begin{equation}\label{Eq10}
\delta\equiv \sqrt{M^2(q^2-\mu^2)-(l+1/2)^2}\  .
\end{equation}

The mandatory existence of a charged particle which respects the
dimensionless weak gravity relation
\begin{equation}\label{Eq11}
{{q}\over{\mu}}>1\
\end{equation}
in the coupled Einstein-Maxwell theory can now be inferred by
substituting $n=0$ in the resonant relaxation spectrum (\ref{Eq9})
and requiring that $\Im\omega_0\leq \pi T_{\text{BH}}$ [see
(\ref{Eq3})] for this fundamental black-hole resonance
\cite{Notedelr,Notestr,Schw}. We have therefore proved the
previously conjectured mass-charge bound (\ref{Eq1}).

{\it Summary.---} The conjectured weak gravity relation $\mu/q<1$
\cite{wgc} in theories describing a U(1) gauge field coupled
consistently to gravity has been the focus of intense research
during the last decade (see \cite{wr1,wr2,wr3,wr4,wr5,wr6,wr7} and
references therein). However, despite the flurry of activity in this
field, the origin of this highly interesting mass-charge relation
has remained somewhat mysterious. In this Letter we have explicitly
shown that the generalized second law of thermodynamics \cite{Bek1},
a physical law which is widely believed to reflect a fundamental
aspect of a self-consistent quantum theory of gravity, may shed much
light on the origins of this highly intriguing bound \cite{Notegu}.
In particular, it has been explicitly proved that the dimensionless
mass-charge upper bound $\mu/q<1$ can be deduced directly from the
self-consistent interplay between quantum theory, thermodynamics,
and gravity.


\bigskip
\noindent
{\bf ACKNOWLEDGMENTS}
\bigskip

This research is supported by the Carmel Science Foundation. I would
like to thank Yael Oren, Arbel M. Ongo, Ayelet B. Lata, and Alona B.
Tea for helpful discussions.


\begin{thebibliography}{99}
\bibitem{Noteelc} This mass-charge relation is certainly true in our world. In particular, the electron is
characterized by the dimensionless mass-to-charge ratio
$m_e/|q_e|\simeq 4.9\times10^{-22}\ll1$.
\bibitem{Noteunit} We shall henceforth use natural units in which
$G=c=\hbar=k_{\text{B}}=1$. Note that in these units $q$ and $m$
have the dimensions of length$^{-1}$.
\bibitem{Notehe} Note, in particular, that the mass-charge inequality (\ref{Eq1}) guarantees
that extremal charged Reissner-Nordstr\"om black holes in the
coupled Einstein-Maxwell theory are quantum mechanically unstable to
the Hawking evaporation process \cite{wgc,Bek1}.
\bibitem{Notegsl} The generalized second law of thermodynamics \cite{Bek1}
states that the total entropy in a coupled black-hole-matter system
is an irreducible physical quantity.
\bibitem{Noteaa} Here $A$ is the black-hole surface area.
\bibitem{Noterelx} Note that the complex nature of the black-hole
resonant frequency $\omega$ reflects the fact that the corresponding
perturbation mode of the composed black-hole-field system decays
over time.
\bibitem{Noterec} Here the characteristic relaxation time of the exponentially
decaying black-hole perturbation mode is given by the relation
$\tau_{\text{relax}}\equiv 1/\Im\omega_0$, where $\omega_0$ is the
fundamental (least damped) resonant frequency of the corresponding
black-hole spacetime.
\bibitem{Noteom} Here $(m,l\geq |m|)$ are the azimuthal and spherical harmonic
indices which characterize the exponentially damped perturbation
modes of the composed black-hole-field system. For brevity, we shall
henceforth omit these integer parameters.
\bibitem{Noters} The radial coordinate $r^*$ in (\ref{Eq6}) is related to the
Schwarzschild areal coordinate $r$ by the differential relation
$dr^*=dr/(1-2M/r+Q^2/r^2)$.
\bibitem{Noteai} This asymptotic boundary condition corresponds to free outgoing
waves (quasinormal resonances) in the large-frequency regime
$\omega^2>\mu^2$, whereas it describes spatially decaying field
configurations (bound-state resonances) in the small-frequency
regime $\omega^2<\mu^2$ \cite{Hodch1,Hodch2,Konf,Ric,Herc}.
\bibitem{Noterpm} Here $r_{\pm}=M\pm(M^2-Q^2)^{1/2}$ are the (outer and inner) horizon
radii of the charged black-hole spacetime.
\bibitem{Notedelr} Note that the dimensionless mass-charge
inequality (\ref{Eq1}) provides a necessary condition for the
relation $\Im\delta=0$. In particular, in this physical regime of
the charged field parameters (with $\delta\in \mathbb{R}$), one
finds a saturation of the universal relaxation bound (\ref{Eq3}) by
the near-extremal ($T_{\text{BH}}\to0$) charged black holes.
\bibitem{Notestr} It is worth noting that one can deduce from Eqs.
(\ref{Eq3}), (\ref{Eq9}), and (\ref{Eq10}) the stronger mass-charge
relation $q/\mu>\sqrt{1+({{l+1/2}\over{M\mu}})^2}$. Thus, the lower
bound on the fundamental ratio $q/\mu$ can be optimized by
minimizing the dimensionless product $M\mu$. At this point, it is
important to stress the fact that the physical phenomenon of quantum
black-hole discharge associated with the Schwinger-type pair
production mechanism in strong electric fields \cite{Schw} sets the
dimensionless quantum upper bound $qQ\ll\mu^2r^2_+/\hbar$
\cite{Hodch2} on the physical parameters of the composed
charged-black-hole-charged-field system. Taking cognizance of the
lower bound (\ref{Eq11}), one obtains the characteristic
inequalities $\mu<q\ll M\mu^2$ in the near-extremal regime. One
therefore finds the characteristic strong inequality $M\mu\gg1$,
which implies $q/\mu>1+O[(M\mu)^{-2}]$.
\bibitem{Notegu} Strictly speaking, we have explicitly shown in this Letter that the
weak gravity relation $\mu/q<1$ can be deduced directly from the
universal relaxation bound (\ref{Eq2}) as applied to the
characteristic complex relaxation spectrum of near-extremal charged
Reissner-Nordstr\"om black holes. It should be remembered, however,
that, as explicitly shown in \cite{Hod1,Hod2}, the universal
relaxation bound can be deduced directly from Bekenstein's
generalized second law of thermodynamics \cite{Bek1}. Thus, one may
regard the weak gravity relation (\ref{Eq1}) as a natural physical
outcome of the generalized second law of thermodynamics.
\end{thebibliography}
\end{document}